\documentclass[12pt]{article}
\usepackage{amsmath}
\usepackage{epsfig}
\usepackage{amsfonts}
\usepackage{amssymb}

\def\beq{\begin{equation}}
\def\eeq{\end{equation}}
\def\bea{\begin{eqnarray}}
\def\eea{\end{eqnarray}}
\def\Tr{{\rm Tr}}
\begin{document}


\begin{titlepage}

\begin{centering}

\vspace*{3cm}

{\Large\bf Towards finding the single-particle content of 
           two-dimensional adjoint QCD}

\vspace*{1.5cm}

{\bf Uwe Trittmann}
\vspace*{0.5cm}

{\sl Department of Physics\\
Otterbein University\\
Westerville, OH 43081, USA}

\vspace*{1cm}

\today

\vspace*{2cm}

{\large Abstract}

\vspace*{1cm}

\end{centering}

The single-particle content of 
two-dimensional adjoint QCD remains elusive due to 
the inability to distinguish single- from multi-particle states.
To find a criterion we compare several approximations to the 
theory.
Starting from the asymptotic theory (no pair production, only 
singular operators), we construct
sets of eigenfunctions in the lowest parton sectors of the 
theory. A perturbative treatment of the omitted
operators is performed. We find that 
multi-particle states are absent if pair-production is disallowed
and hints for a double Regge trajectory of single-particle states.
We discuss the structure of the eigensystem of the theory, and present
the reason for the fact that bosonic single-particle states  
do not form multi-particle states. 
\vspace{0.5cm}


\end{titlepage}
\newpage


\section{Introduction}

Two-dimensional Yang-Mills theory
coupled to fermions in the adjoint representation, QCD$_{2A}$, has 
been discussed extensively in the literature
\cite{Kutasov94, DalleyKlebanov, BDK, GHK, Katz}, due to its many 
interesting features (see, e.g. \cite{LesHouches}). 
However, its single-particle
spectrum remains elusive, largely because there is no clear
criterion to help purge the theory of its multi-particle content.
Recently, QCD$_{2A}$ has been numerically solved as a fermion theory.
In \cite{Katz}, the authors use an idea from holography, namely 
that the theory is a trivial CFT in the UV limit. Therefore a decoupling
between the low-lying spectrum and the high scaling-dimension quasi-primary
operators ensues. A basis of these operators is constructed and
cut off at a maximal (scaling) dimension. Good agreement is found with 
previous DLCQ results \cite{BDK,GHK,UT}. 
While \cite{Katz} furnishes an important contribution to the ongoing
debate over the single-particle content of QCD$_{2A}$, 
the disappointing conclusion is that also this approach is riddled with 
multi-particle states. 

We present some progress on teasing out the true 
(single-particle) content of the theory described in more detail in 
Sec.~\ref{SecTheTheory}. We start in Sec.~\ref{SecNoPairs} by considering 
the asymptotic approach of \cite{Kutasov94}, in which the theory is solved for 
high excitation numbers, {\em i.e.}~in a regime where parton number is conserved. 
We then explore the impact of non-singular
interactions on the spectrum in 
Sec.~\ref{SecNonSingular}. Sec.~\ref{SecPairs}
is devoted to the role of the pair-production operators and the emergence
of multi-particle states. Finally, 
we take a look at the implications of bosonization
in Sec.~\ref{SecCurrents} and conclude.

\section{The Spectrum of QCD$_{2A}$}
\label{SecTheTheory}

Adjoint QCD$_{2}$ is based on the following Lagrangian
in light-cone coordinates $x^\pm=(x^0\pm x^1)/\sqrt{2}$, where $x^+$ plays the
role of a time
\beq
{\cal L}=Tr[-\frac{1}{4g^2}F_{\mu\nu}F^{\mu\nu}+
i\bar{\Psi}\gamma_{\mu}D^{\mu}\Psi],
\eeq
where $\Psi=2^{-1/4}({\psi \atop \chi})$, 
with $\psi$ and $\chi$ 
being $N_c\times N_f$ matrices. The field strength is
$F_{\mu\nu}=\partial_{\mu}A_{\nu}-\partial_{\nu}A_{\mu}+i[A_{\mu},A_{\nu}]$,
and the covariant derivative is defined as $D_{\mu}=\partial_{\mu}
+i[A_{\mu},\cdot]$. 
Working in the light-cone gauge, $A^+=0$, is consistent if  
the fermionic zero modes are omitted. 
The left-moving fermions can be integrated out, and the light-cone 
momentum $P^+$ and Hamiltonian $P^-$ can be written in terms of the Fourier
oscillation modes of the right-moving fermion only \cite{BDK,Anton}.
Once the theory is formulated in terms of independent degrees of freedom,
we can quantize it by imposing canonical anti-commutation relations at equal 
light-cone times $x^+$ 
\beq\label{PhiCR}
\left\{\psi_{ij}(x^{-}), \psi_{kl}(y^{-})\right\} = \frac{1}{2}\,
\delta(x^{-}-y^{-})\bigl ( \delta_{il} \delta_{jk}-
{1\over N}\delta_{ij} \delta_{kl}\bigr).
\eeq
One uses the usual decomposition of the fields in 
terms of fermion operators
\beq\label{PhiExpansion}
\psi_{ij}(x^-) = {1\over 2\sqrt\pi} \int_{0}^{\infty} dk^{+}
\left(b_{ij}(k^{+}){\rm e}^{-ik^{+}x^{-}} +
b_{ji}^{\dagger}(k^{+}){\rm e}^{ik^{+}x^{-}}\right )\ ,
\eeq
with anti-commutation relations following from Eq.~(\ref{PhiCR})
\beq\label{Commy}
\{b_{ij}(k^{+}), b_{lk}^{\dagger}(p^{+})\} =
\delta(k^{+} - {p}^{+})
(\delta_{il} \delta_{jk}-{1\over N}\delta_{ij} \delta_{kl})\,\,
\eeq
to write the operators in terms of oscillators  
\begin{eqnarray}
P^+ &=& \int_{0}^{\infty} dk\ k\, b_{ij}^{\dagger}(k)b_{ij}(k)\ ,\\
P^{-} &=& {m^2\over 2}\, \int_{0}^{\infty}\label{P_BDK}
{dk\over k} b_{ij}^{\dagger}(k)
b_{ij}(k) +{g^2 N\over \pi} \int_{0}^{\infty} {dk\over k}\
C(k) b_{ij}^{\dagger}(k)b_{ij}(k) \\
&&+ {g^2\over 2\pi} \int_{0}^{\infty} dk_{1} dk_{2} dk_{3} dk_{4}
\biggl\{ B(k_i) \delta(k_{1} + k_{2} +k_{3} -k_{4})\nonumber \\
&&\qquad\qquad\times(b_{kj}^{\dagger}(k_{4})b_{kl}(k_{1})b_{li}(k_{2})
b_{ij}(k_{3})-
b_{kj}^{\dagger}(k_{1})b_{jl}^{\dagger}(k_{2})
b_{li}^{\dagger}(k_{3})b_{ki}(k_{4})) \nonumber\\
&&\qquad + A(k_i) \delta (k_{1}+k_{2}-k_{3}-k_{4})
b_{kj}^{\dagger}(k_{3})b_{ji}^{\dagger}(k_{4})b_{kl}(k_{1})b_{li}(k_{2})
 \nonumber\\
&&\qquad + \frac{1}{2} D(k_i) \delta (k_{1}+k_{2}-k_{3}-k_{4})
b_{ij}^{\dagger}(k_{3})b_{kl}^{\dagger}(k_{4})b_{il}(k_{1})b_{kj}(k_{2})
\biggl\} \nonumber
\end{eqnarray}
with
\begin{eqnarray}
A(k_i)&=& {1\over (k_{4}-k_{2})^2 } -
{1\over (k_{1}+k_{2})^2}\ , \label {A} \label{EqnA}\\
B(k_i)&=& {1\over (k_{2}+k_{3})^2 } - {1\over (k_{1}+k_{2})^2 }, \label {B}\\
C(k)&=& \int_{0}^{k} dp \,\,{k\over (p-k)^2},\\ 
D(k_i)&=& \frac{1}{(k_{1}-k_{4})^2} - \frac{1}{(k_{2}-k_{4})^2},
\end{eqnarray}
where the trace-splitting term $D(k_i)$ can be omitted at large $N_c$, and 
the trace-joining term is proportional to $B(k_i)$. 
The structure of the Hamiltonian $P^-$ displayed in Eq.~(\ref{P_BDK}) is 
\beq\label{StructureOfHamiltonian}
P^-= P^-_{m}+P^-_{ren}+ P^-_{PC,s}+P^-_{PC,r}+P^-_{PV} + P^-_{finiteN} .
\eeq
Obviously, the mass term $P^-_{m}$ is dropped in the massless theory, yet the 
renormalization operator $P^-_{ren}$ needs to be included. 
Interactions that violate parton number, $P^-_{PV}$, couple blocks of 
different parton number, whereas parton-number conserving interactions 
$P^-_{PC}$ are block diagonal, and may include singular($s$) or regular($r$) 
functions of the parton momenta.

If one considers large excitation numbers, parton-number violating operators 
proportional to $B(k_i)$ 
can be neglected and the mass of the fermions becomes 
irrelevant \cite{Kutasov94}. 
We will refer to the resulting approximation as the {\em asymptotic theory}:
we retain the most singular terms in the interaction
only, and additionally use the approximation
\beq\label{approx}
\int^1_0\frac{dy}{(x-y)^2}\phi(y) \approx 
\int^\infty_{-\infty}\frac{dy}{(x-y)^2}\phi(y), 
\eeq
because for the highly excited states the integral is dominated by the 
interval around $x=y$, associated with the long-range Coulomb-type force.  
Thus, the asymptotic theory is split into decoupled sectors with 
fixed parton numbers subject to 't Hooft-like equations
\beq
M^2\phi_r(x_1,\ldots,x_r)=-\sum_{i=1}^r(-1)^{(r+1)(i+1)}
\int_{-\infty}^{\infty}\frac{\phi_r(y,x_i+x_{i+1}-y,
x_{i+2},\ldots,x_{i+r-1})}{(x_i-y)^2}dy,
\label{TheEquation}
\eeq
where the wavefunctions $\phi_r$ distribute momentum in the 
states of definite parton number $r$
\beq\label{TheStates}
|\Phi_r\rangle=\left(\prod_{j=1}^r\int_0^1 dx_j\right)\,
\delta(1-\sum^r_{i=1} x_i)
\phi_r(x_1,x_2,\ldots,x_r) \frac{1}{N_c^{r/2}}
Tr[b(-x_1)\cdots b(-x_r)]|0\rangle.
\eeq
The $x_i$ are momentum fractions with $\sum_i x_i=1$, and the total 
momentum has been set to unity.
The number of partons $r$ is even (odd) for bosonic (fermionic) states. 

A complete set of solutions of Eq.~(\ref{TheEquation})
remains elusive, while Ref.~\cite{Kutasov94} displays what looks like 
half of the bosonic eigenfunctions, {\em i.e.}~even-$r$ eigenfunctions with 
eigenvalues $(-1)^{r/2+1}$ under the theory's $Z_2$ orientation symmetry
\beq\label{Tsymmetry}
{\cal T}: b_{ij} \rightarrow b_{ji}.
\eeq 
In these sectors, the eigenfunctions listed in \cite{Kutasov94}
have eigenvalues
\beq\label{EigenValues}
M^2_{n_1,\ldots, n_k}= 2g^2N\pi^2(n_1+n_2+\cdots+n_k),
\eeq
where the excitation numbers $n_i$ are 
even and their sum is much larger than $k\equiv r/2$.
This implies an exponentially 
growing density of states, and points towards the existence of a 
Hagedorn transition of the theory at high temperatures.
Eq.~(\ref{EigenValues}) 
suggests a $r/2$-dimensional manifold of solutions in the $r$-parton sector.
However, the $r-1$ relative momenta of the sector lead one
to expect $r-1$ quantum numbers. Incidentally, 
an $r/2$-dimensional manifold of solutions
makes it hard to think of a generalization to 
the fermionic (odd $r$) sectors of the theory. 
The functions displayed in \cite{Kutasov94} are therefore likely
particular solutions; the general solutions should exhibit
additional excitation numbers.

The clear separation of the eigenvalues, 
Eq.~(\ref{EigenValues}), does not guarantee that these are single-particle
solutions. We know from \cite{GHK} that exact and approximate multi-particle
states exist in the single-trace sector of the theory, so that single-particle
states cannot be identified with single-trace states. 
The problem is compounded by the approximations made.
While omitting the non-singular terms in the interaction and discarding 
parton-changing operators can be justified on physical grounds, approximating 
the integral as in Eq.~(\ref{approx}) implies unphysical effects which 
paradoxically make the solutions simpler.
Furthermore, the correct generalization 
of 't Hooft's approximations \cite{tHooft}
to higher parton sectors is a restriction of the Hilbert space from
the naive $[0,1]^r$ hypercube to a $(r-1)$-simplex, which takes 
up $1/r!$ of the former's volume, see Appendix. 
We expect fewer 
linearly independent eigensolutions on the simplex than on the hypercube.

In fact, multi-particle states (identified by their threshold masses) 
are absent altogether in the
asymptotic theory. A quick DLCQ calculation traces this behavior 
back to the absence of parton-number violation.
This means that a method to distinguish single- from
multi-particle states cannot emerge from the asymptotic theory alone. 
Identifying threshold mass values as in \cite{GHK} 
is not going to be good enough either: the alleged multi-particle
states fulfill a single-particle integral equation\footnote{I am 
grateful to D.G.~Robertson for pointing this out.}. 
On the other hand, one knows from the bosonized theory
that states absent in the adjoint and
identity block of the current algebra are true multi-particle
states \cite{KutasovSchwimmer}, 
and one can study them. The opposite is not 
true, and one has to learn how to identify the single-particle states
in these current blocks.
Unfortunately, it is unlikely that approximate solutions 
{\'a la} 't Hooft \cite{tHooft} and Kutasov \cite{Kutasov94} 
exist in the bosonized theory, because
bosonization implies parton-number violation.

The eigenvalue problem 
at hand
is equivalent to an integral equation which is completely specified 
{\em ab ovo}. As such, Eq.~(\ref{TheEquation}) 
implies that its solutions fulfill several 
constraints: 
the (pseudo-)cyclicity of the wavefunction
\beq\label{Cyclicity}
\phi_r(x_1,x_2,\ldots,x_r)=(-1)^{r+1}\phi_r(x_2,x_3\ldots,x_r,x_1),
\eeq
since the fermions are real, and 
the constraint  
\beq\label{ZeroBC}
\phi_n(0,x_2, \ldots, x_n)=0,
\eeq
necessary to secure hermiticity of the Hamiltonian (only) 
in the presence of a mass term\footnote{The apparent vanishing of the 
${\cal T}=(-1)^{r/2}[(-1)^{(r-1)/2}]$ wavefunctions for even [odd] $r$ 
at the ends of the intervals, see Fig.~\ref{fig3p}, is not due to a 
boundary condition but accidental; the computer code happens to 
choose $|1,1,\ldots, K-r-1\rangle$ as first, and 
$|K/r,K/r,\ldots, K/r\rangle$ (or similar) as last basis state. At these 
points the eigenfunctions vanish due to symmetry constraints.}. 
In the case of the 't Hooft model \cite{tHooft}, this amounts to a 
``boundary condition'' in the sense that the values of the wavefunction
are specified at the endpoints of the interval. 
We find it advantageous to realize (and in some sense relax) the latter 
constraint by replacing it with the condition
\beq\label{AltZeroBC}
\phi_n(x_1,x_2, \ldots, x_n)=\pm \phi_n(1-x_1,1-x_2, \ldots, 1-x_n),
\eeq
which allows for a natural interpretation of the massless (massive) 
theory's solutions as (anti-)periodic functions. 
Of course, all constraints are fixed by the form of the integral equation, 
and cannot 
to be confused with the conditions specified to solve a differential equation. 
For instance, hermiticity is given, the vanishing of the 
wavefunctions follows. 

Solutions of definite $\cal T$-symmetry, Eq.~(\ref{Tsymmetry}),
fulfill an additional condition, which means that the 
{wavefunctions} have different support. Namely, some combination 
of creation operators might not exist in one symmetry sector. 
For example, in the four-parton sector a constraint arises because  
states like $Tr[b(-x)b(-x)b(-y)b(-y)]|0\rangle$ are $\cal T$ even.
Analogous requirements exist in other sectors\footnote{For 
the first few parton sectors they are: 
$\phi_{3-}(x,x,y)=0$, 
$\phi_{4+}(x,y,x,y)=0$, $\phi_{4-}(x,x,y,y)=0$,
$\phi_{5+}(x,y,y,x,z)=0$, 
$\phi_{6+}(x,y,y,x,z,z)=0$, $\phi_{6-}(x,y,z,w,z,y)=0$, and cyclic.}, except 
for the fermionic ${\cal T}=(-1)^{(r+1)/2}$ sectors\footnote{Incidentally, 
these sectors sport a massless state when the above approximations are used.}.

\section{Solving the asymptotic eigenvalue problem}
\label{SecNoPairs}

We can solve the Kutasov integral equation (4.10) of 
Ref.~\cite{Kutasov94} algebraically by using the following ansatz for 
the wavefunctions
\beq\label{Ansatz}
|n_1, n_2,\ldots n_{r-1} \rangle \doteq \prod^{r-1}_j e^{i \pi n_j x_j}
=\phi_r(x_1,x_2,\ldots,x_r),
\eeq
where $r$ is the number of partons, $x_r=1-\sum_j^{r-1} x_j$. 
Note that we have $r-1$ excitation 
numbers $n_i$, 
as expected from $r-1$ relative momenta in the $r$ parton sector.
The $r=2$ version solves 
the 't Hooft equation 
\beq\label{r2tHooft}
\frac{M^2}{g^2 N_c} e^{i\pi n x} = -\int^{\infty}_{-\infty} 
\frac{dy}{(x-y)^2} e^{i\pi n y} = \pi^2 |n| e^{i\pi n x}. 
\eeq
In other words, we use the single-particle states of a Hamiltonian
appropriate for the problem to construct a Fock basis,  
in the spirit of Ref.~\cite{Pauli84}.  
These single-particle states are two-parton states, 
and they constitute an orthonormal basis on the
interval $[0,1]$. However, the multi-parton states live in a restricted
Hilbert space because the total momentum is fixed, see Appendix. 
Clearly, Eq.~(\ref{r2tHooft}) is insensitive to the sign of $n$. Hence, we 
admit positive and negative 
excitation numbers: 
$n_i \in 2 {\mathbb Z}$ or $2 {\mathbb Z}+1$. 

There is a rather elegant solution to the eigenvalue 
problem, Eq.~(\ref{TheEquation}), based on the observation
that the solutions of the adjoint 't Hooft problem have to be (anti-)cyclic,
{\em cf.}~Eq.~(\ref{Cyclicity}).
By introducing the cyclic permutation operator
\[
{\cal C}: (x_1,x_2,\ldots,x_r)\rightarrow(x_2,x_3,\ldots,x_r,x_1),
\]
we can construct the solution to the asymptotic 
adjoint 't Hooft problem by symmetrizing our ansatz
\beq\label{AlgebraicSolution}
|n_1,n_2, \ldots n_{r-1}\rangle_{sym}\equiv \frac{1}{\sqrt{r}}
\sum_{k=1}^r (-1)^{(r-1)(k-1)}{\cal C}^{k-1}|n_1,n_2, \ldots n_{r-1}\rangle,
\eeq
where ${\cal C}^{0}=\boldmath{1}$.
This furnishes a general asymptotic solution of adjoint QCD$_2$. 
It is not hard to show that the eigenvalues are
\beq\label{AnsatzMasses}
M^2 = g^2 N \pi^2\sum_{k=1}^{r}|n^{(k-1)}_1-n^{(k-1)}_2|=
g^2 N \pi^2\sum_{k=1}^{r}|n^{(k-1)}_1|,
\eeq
where $n_i^{(k)}$ is the excitation number associated with the 
$i$th momentum fraction of the $k$th cyclic 
permutation, e.g.~${\cal C}^2|n_1,n_2,n_3\rangle$ yields 
$|n^{(2)}_1-n^{(2)}_2|=|n_3-n_2|+|-n_2|$. 
This could be a useful method for similar integral equations,
like the one associated with adjoint Dirac fermions recently tackled in
\cite{GHKSS}. 

First, let us clean up the spectrum by using
the orientation symmetry $\cal T$ of the Hamiltonian, Eq.~(\ref{Tsymmetry}).
Note that both symmetry operators act a bit awkwardly on the basis states,
as they are naturally defined with $r$ variables, but actually live in a
$(r-1)$-dimensional space
\bea
{\cal C}&:& |n_1,n_2,\ldots,n_{r-1}\rangle 
\rightarrow (-1)^{n_{r-1}} |-n_{r-1},n_1-n_{r-1},n_2-n_{r-1},\ldots, n_{r-2}-n_{r-1} \rangle, \nonumber
\\
{\cal T}&:& |n_1,n_2,\ldots,n_{r-1}\rangle 
\rightarrow (-1)^{n_1} |-n_1,n_{r-1}-n_1,n_{r-2}-n_1,\ldots, n_2-n_1 \rangle,
\label{CTaction}
\eea
While $[{\cal C}^k,{\cal T}]\neq 0$, 
except for trivial cases, we have
\[
\left[\sum_{k=1}^r {\cal C}^{k-1},{\cal T}\right]=0,
\]
and, by construction, $[(-1)^{k(r-1)}{\cal C}^{k},P^-]=0$, 
so we can classify the eigenstates according to their eigenvalues 
$M^2$ and $T$. To fulfill the integral (eigenvalue) equation, one has to 
choose one specific $\cal C$ eigenvalue. 

%
\begin{figure}
\centerline{
\psfig{file=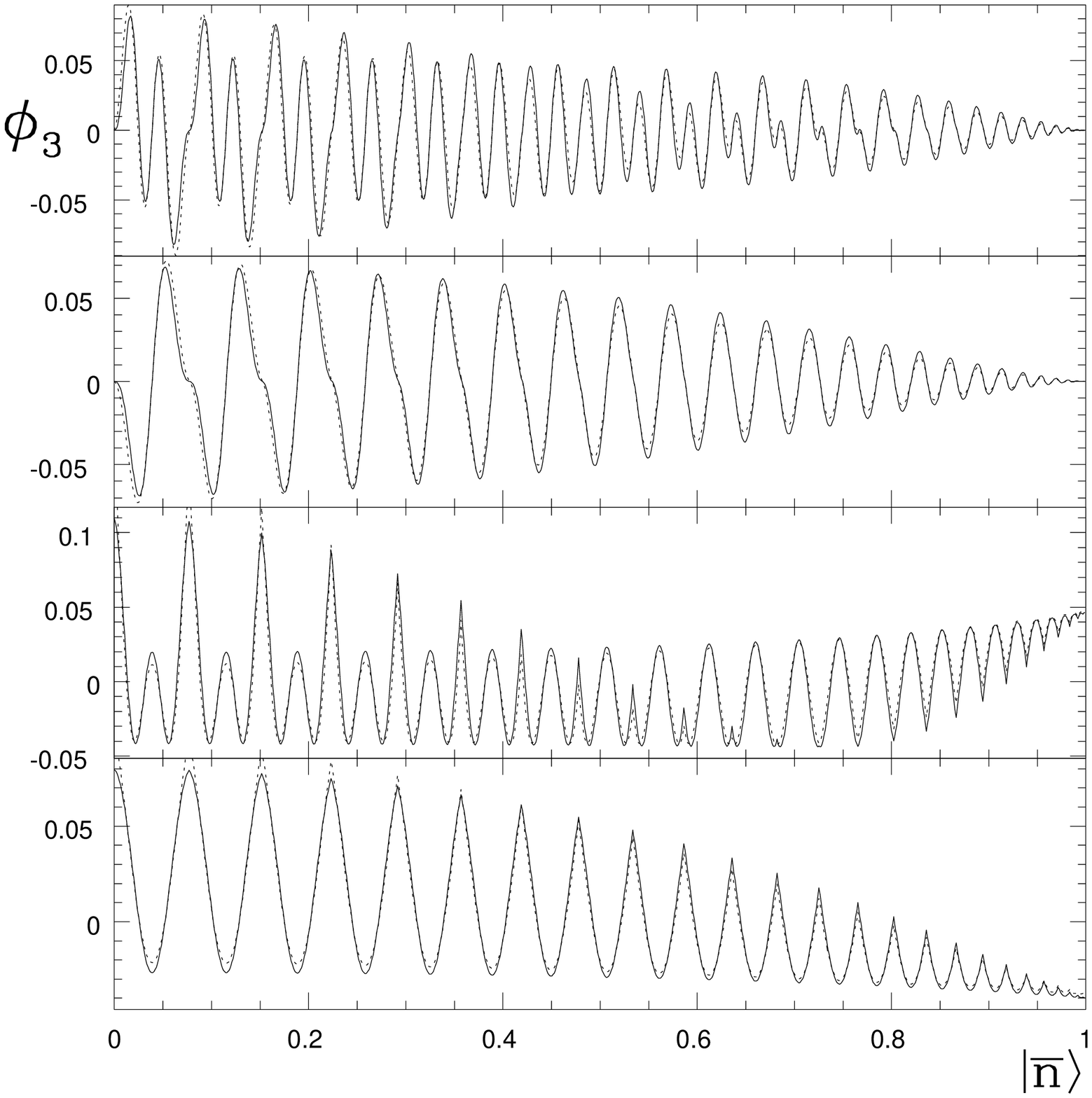,width=7.8cm}
\psfig{file=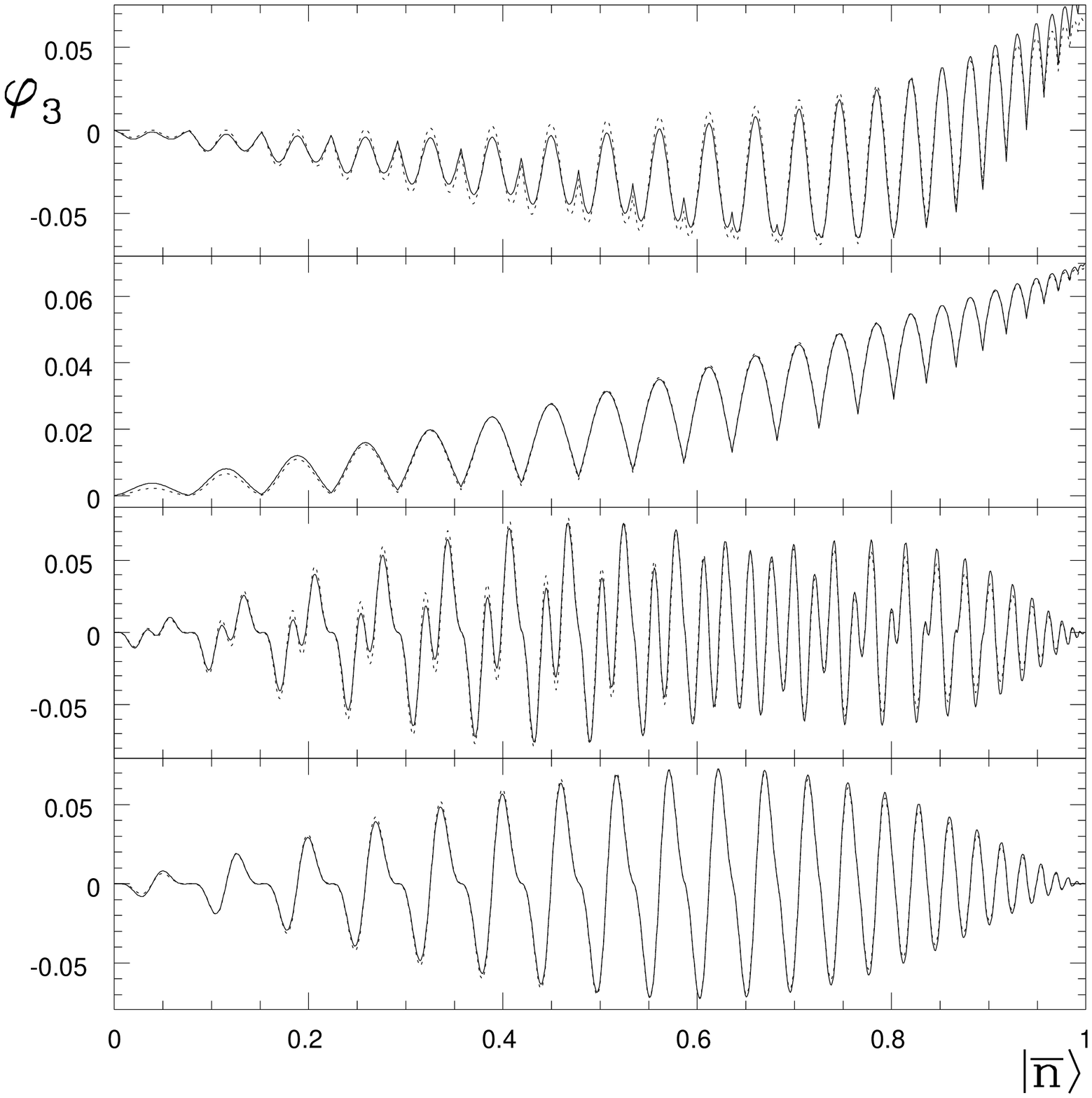,width=7.8cm}
}
\caption{DLCQ eigenfunctions (solid lines) and asymptotic 
wavefunctions (dashed lines) 
of the theory without pair-production and non-singular terms.  
(a) Left: The lowest two three-parton eigenfunctions in the 
$\cal T$- even and -odd sectors (from bottom); $K=151$ in the DLCQ 
calculation with massless fermions. 
(b) Right: Same for massive theory.  
\label{fig3p}}
\end{figure}
%

As a cross-check of our ansatz, we will compare to numerical wavefunctions
generated by a DLCQ algorithm. A further check is provided
by the solutions listed in Ref.~\cite{Kutasov94}, which can be emulated within
DLCQ by choosing a large fermion mass, which enforces the constraint,
Eq.~(\ref{ZeroBC}) or (\ref{AltZeroBC}), {\em vulgo} the 
vanishing of the wavefunction at the boundaries. 
We will refer to the latter solutions as massive parton solutions.
We need to construct a complete orthonormal basis of the physical Hilbert 
space from the ansatz, Eq.~(\ref{AlgebraicSolution}). 
We'll work out the solutions 
in the first few sectors, and develop a general algorithm for the others. 

At $r=2$ we have 
${\cal C}|n\rangle={\cal T}|n\rangle=(-1)^n|-n\rangle$, hence
\beq\label{phi2}
\phi_2=e^{i\pi nx}-(-1)^n e^{-i\pi nx}.
\eeq
Thus both sines with even $n$ and cosines with odd $n$ fulfill the 
integral equation, the cyclicity condition, and are states of definite $T$. 
Physics determines which functions to pick: massive partons require
$\phi_2(0)=0$ or $\phi_2(x)=-\phi_2(1-x)$, whereas a massless theory requires  
odd $n$ cosines, {\em i.e.} $\phi_2(x)=\phi_2(1-x)$, 
since a massless bound-state with a 
constant wavefunction exists in the limit $N_f\rightarrow 1$. 
We clean up the notation for the generic case, rewriting 
Eq.~(\ref{phi2}) as
\[
|\phi_2,n; \bar{M}^2=|n|\rangle_-=
|n\rangle -(-1)^n |-n\rangle,
\]
where $\bar{M}^2=M^2/{g^2 N \pi^2}$, and the minus index signifies that only
the wavefunction odd
under the ${\cal T}$ operation exists.


In the three-parton sectors, $r=3$, 
we find that both excitation numbers have to be even, 
because a massless bound state with constant 
wavefunction exists. 
For massive
partons, no massless state exists, but the eigensolutions are again from 
the even-even
$\{|ee\rangle\}$ 
sector, cf.~Figure \ref{fig3p}. The reason is that 
the ${\cal C,T}$ operators permute excitation numbers, 
{\em cf.}~Eqs.~(\ref{CTaction}), generating 
combinations like $n-m$ which are even for $n,m$ odd.

The wavefunctions of definite ${\cal C,T}$ symmetry are
\bea
\label{r3CTstates}
&&\!\!\!\!\!\!\!\!\!\!\!\! |\phi_3,n,m; \bar{M}^2=|n-m|+|n|+|m|\rangle_{\pm}=\\
&&\qquad\qquad\qquad
|n,m\rangle +(-1)^m |-m,n-m\rangle+(-1)^n |m-n,-n\rangle
\nonumber\\  
&&\qquad\qquad\qquad\qquad
\pm\left[(-1)^n |-n,m-n\rangle 
+ |m,n\rangle+(-1)^m |n-m,-m\rangle  
\right],
\nonumber
\eea 
which are symmetric $(+)$ or antisymmetric $(-)$ under reversal of momentum 
fractions.
Note that some solutions do not exist in the ${\cal T}$-odd sector, e.g.
$|\phi_3,n=-2,m=0; \bar{M}^2=4\rangle_-=0$. The 
massless solution has a constant wavefunction with $n=m=0$.

Note that the states, Eq.~(\ref{r3CTstates}), are not eigenfunctions of the 
Hamiltonian, because they do not fulfill Eq.~(\ref{AltZeroBC}).
In order to create (anti-)symmetric wavefunctions we must
combine positive and negative frequency solutions.
This is natural, since the fermions are real, 
and hence the eigenfunctions can be chosen to be real. For instance
\bea\label{AbstractPhi3}
Re|\phi_3,n,m\rangle
&=&
\cos (\pi n x_1 + \pi m x_2)
+(-1)^m\cos(-\pi m x_1+\pi (n-m) x_2)\nonumber\\
&&+(-1)^n\cos(\pi (m-n) x_1-\pi n x_2)
\pm (\bar{n}_1 \leftrightarrow \bar{n}_2),\nonumber 
\eea
where $\bar{n}_i$ is the excitation number associated with $x_i$ in a term,
e.g. $\bar{n}_1=-m$ in the $(-1)^m$ term, 
which have to be permuted to obtain a state of definite symmetry under 
reversal of momentum fractions due to the ${\cal T}$ symmetry; the
$(-1)^{n_i}$ 
factors remain unchanged. Note the disappearance of the $\pm$ sign:
the real wavefunctions are all symmetric under momentum fraction reversal;
the antisymmetric functions are identically zero.
We can transcribe the wavefunction into $(x_1,x_2,x_3)$ notation 
to obtain an expression manifestly symmetrized in the momentum fractions
\bea
\phi_{3+}^{(n,m)}(x_1,x_2,x_3)&=&
\sum_{i=1}^{3} \cos (\pi n x_i + \pi m x_{i+1})  
+ (n \leftrightarrow m),
\eea 
The functions with the lowest excitation numbers are a decent fit to the lowest
(DLCQ) eigenfunctions, {\em i.e.} $|1\rangle=\phi_{3+}^{(0,0)}=const.$,
$|2\rangle=\phi_{3+}^{(2,0)}=\phi_{3+}^{(2,2)}$, 
$|3\rangle=\phi_{3+}^{(2,-2)}=\phi_{3+}^{(4,2)}$, 
$|4\rangle=\phi_{3+}^{(4,0)}$,
$|5\rangle=\phi_{3+}^{(6,2)}$, see Fig.~\ref{fig3p}.
From Eq.~(\ref{r3CTstates}) it is clear that 
$\phi_{3\pm}^{(n,m)}=\pm\phi_{3\pm}^{(m,n)}$,
and $\phi_{3+}^{(n,m)}=\phi_{3+}^{(-n,-m)}$, but note that in general 
distinct sets of excitation numbers do not result in distinct wavefunctions.

The odd $T$ solutions are the imaginary part of the general
wavefunction
\bea\label{phi3minus}
\phi_{3-}^{(n,m)}(x_1,x_2,x_3)
&=&
\sum_{i=1}^{3} \sin (\pi n x_i + \pi m x_{i+1})  - (n \leftrightarrow m).
\eea
Again, the functions with the lowest excitation numbers 
are a decent fit to the lowest (DLCQ)
eigenfunctions, {\em i.e.}~$|1\rangle=\phi_{3-}^{(4,2)}$,
$|2\rangle=\phi_{3-}^{(6,2)}$, 
$|3\rangle=\phi_{3-}^{(8,2)}$, 
$|4\rangle=\phi_{3-}^{(8,4)}$, see Fig.~\ref{fig3p}. Note that 
$\phi_{3-}^{(n,n)}=\phi_{3-}^{(n,0)}=0$ and 
$\phi_{3-}^{(n,m)}=-\phi_{3-}^{(-n,-m)}$. 

The massive parton solutions $\varphi_3$ 
are well-described by the same 
formulae in the opposite ${\cal T}$-sector, {\em i.e.}
\[
\varphi_{3\pm}^{(n,m)}(x_1,x_2,x_3)=\phi_{3\mp}^{(n,m)}(x_1,x_2,x_3),
\]
with the excitation numbers of the lowest eigenfunctions being 
$(2,0),(4,0),(6,2),(6,0)$ 
and  $(6,2),(8,2),(10,4),(10,2)$, 
in the ${\cal T}$-even and ${\cal T}$-odd sectors, respectively. 
In the latter sector many functions are identically zero due to 
$\varphi_{3-}^{(n,m)} = -\varphi_{3-}^{(n, n-m)}$. 

As a more stringent test of our ansatz we expanded the numerical solutions into
the complete set of functions just derived, and checked 
that the coefficients of the expansion fall off fast. 
Note, though, that we are comparing numerical eigensolutions
of the true, amputated\footnote{Correct integral 
boundaries are used, but parton-number 
violating terms have been chopped off.} Hamiltonian, with analytic 
eigensolutions of the asymptotic Hamiltonian.
Surprisingly, the eigenfunctions are perfectly reproduced
with only a few non-vanishing coefficients, while the eigenvalues are
off. For instance, at $r=3$ 
ten basis states produce overlaps of 
larger than 99.5\% with the first few eigenfunctions 
in the sector with the massless state, and the overlaps with the tenth 
function are in the permille range.  
The conclusion is that for low excitation numbers the mass renormalization
term and the true integral limits are important to obtain the correct
eigenvalues, whereas the symmetries of the system (cyclicity of the integral
equation, orthogonality constraints of the physical Hilbert space) are so
stringent that, assuming sinusoidal functions, there is very little leeway
to choose the eigenfunctions, so they are basically fixed. 



At $r=4$, the states of 
definite ${\cal C,T}$ symmetry are 
{\small
\bea\label{phi4}
&&\!\!\!\!\!\!\! |\phi_4,n,m,l; 
\bar{M}^2=|n-m|+|n|+|l|+|m-l|\rangle_{\pm}=\\
&&
|n,m,l\rangle -(-1)^l |-l,n-l,m-l\rangle+(-1)^m |l-m,-m,n-m\rangle
-(-1)^n |m-n,l-n,-n\rangle
\nonumber\\  
&&\!\!\!\!\!\!\!\!\!\pm\left[(-1)^n |-n,l-n,m-n\rangle 
- |l,m,n\rangle+(-1)^l |m-l,n-l,-l\rangle-(-1)^m |n-m,-m,l-m\rangle  
\right]\nonumber
\eea 
}
Due to the intricate way the excitation numbers are linked to the mass of 
the bound state, Eq.~(\ref{AnsatzMasses}),
states with distinct sets of excitation numbers may have identical masses. 
The (orthogonal) eigenstates of the
Hamiltonian are thus linear combinations of these states.
For instance, the lightest states, with ${\bar M}^2=2(|n_1|+|n_2|)$ stem 
from the combination
\bea\label{KutLinComb}
\phi_{4-}(x_1,x_2,x_3,x_4)&\doteq&
|\phi_4,n_1,0,n_2\rangle-|\phi_4,n_1,0,-n_2\rangle
+|\phi_4,n_1,n_1-n_2,-n_2\rangle,
\eea 
which is the subset of solutions displayed
as wavefunctions $\phi_4(x_1,x_2,x_3,x_4)$ in Ref.~\cite{Kutasov94}, 
Eq.~(4.13). 
We find empirically 
that the excitation numbers are all even and that there are other solutions
not describable by Eq.~(\ref{KutLinComb}). 

In summary, we see that the symmetrization of states with $r-1$ excitation 
numbers yields a surprisingly simple solution for adjoint QCD$_2$, 
and is in agreement with previous results, which turn out to be special cases
of the general eigensolutions presented here.

\section{The impact of non-singular operators}
\label{SecNonSingular}

%
\begin{figure}[h]
\centerline{
\psfig{file=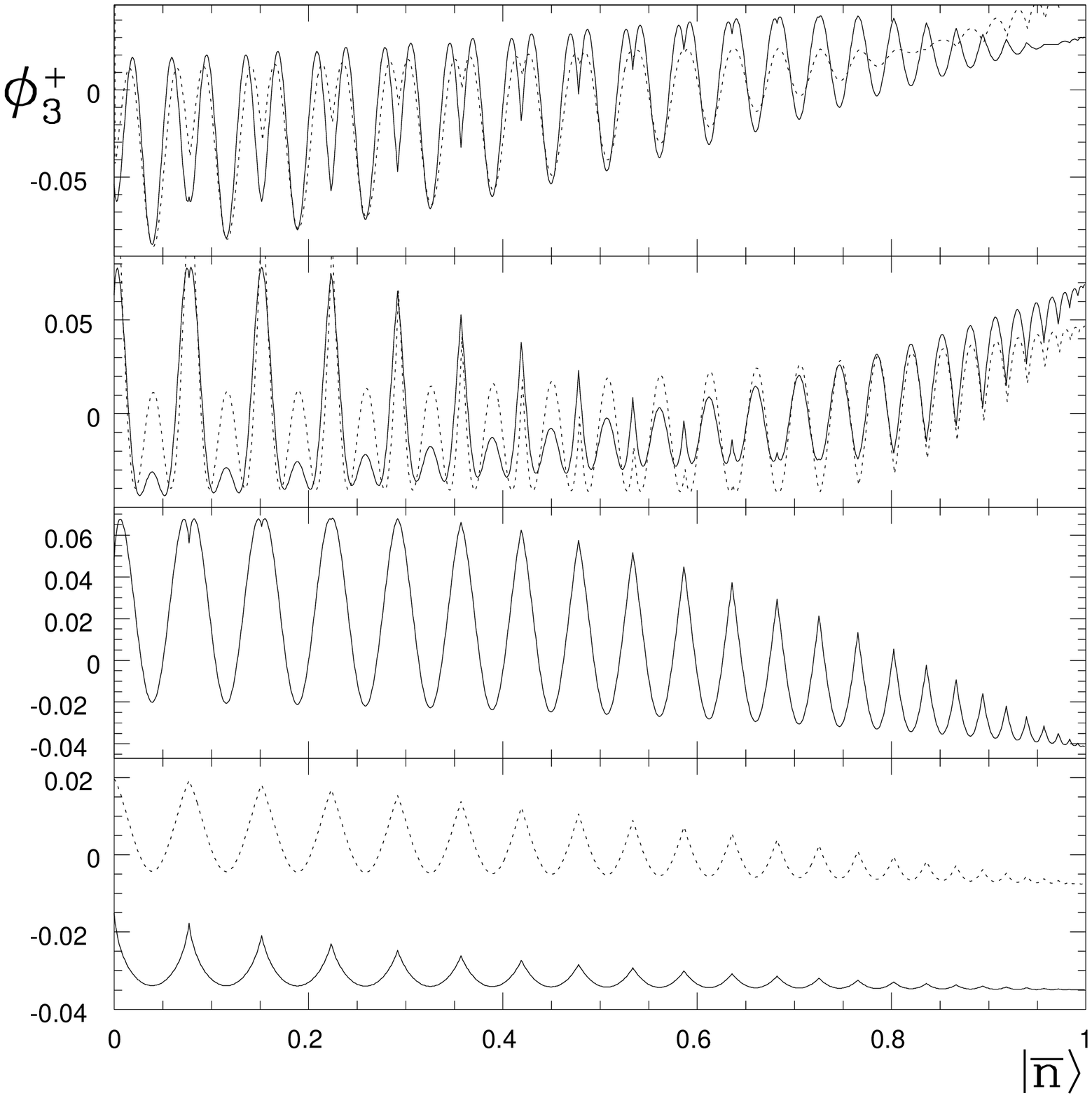,width=7.8cm}
\psfig{file=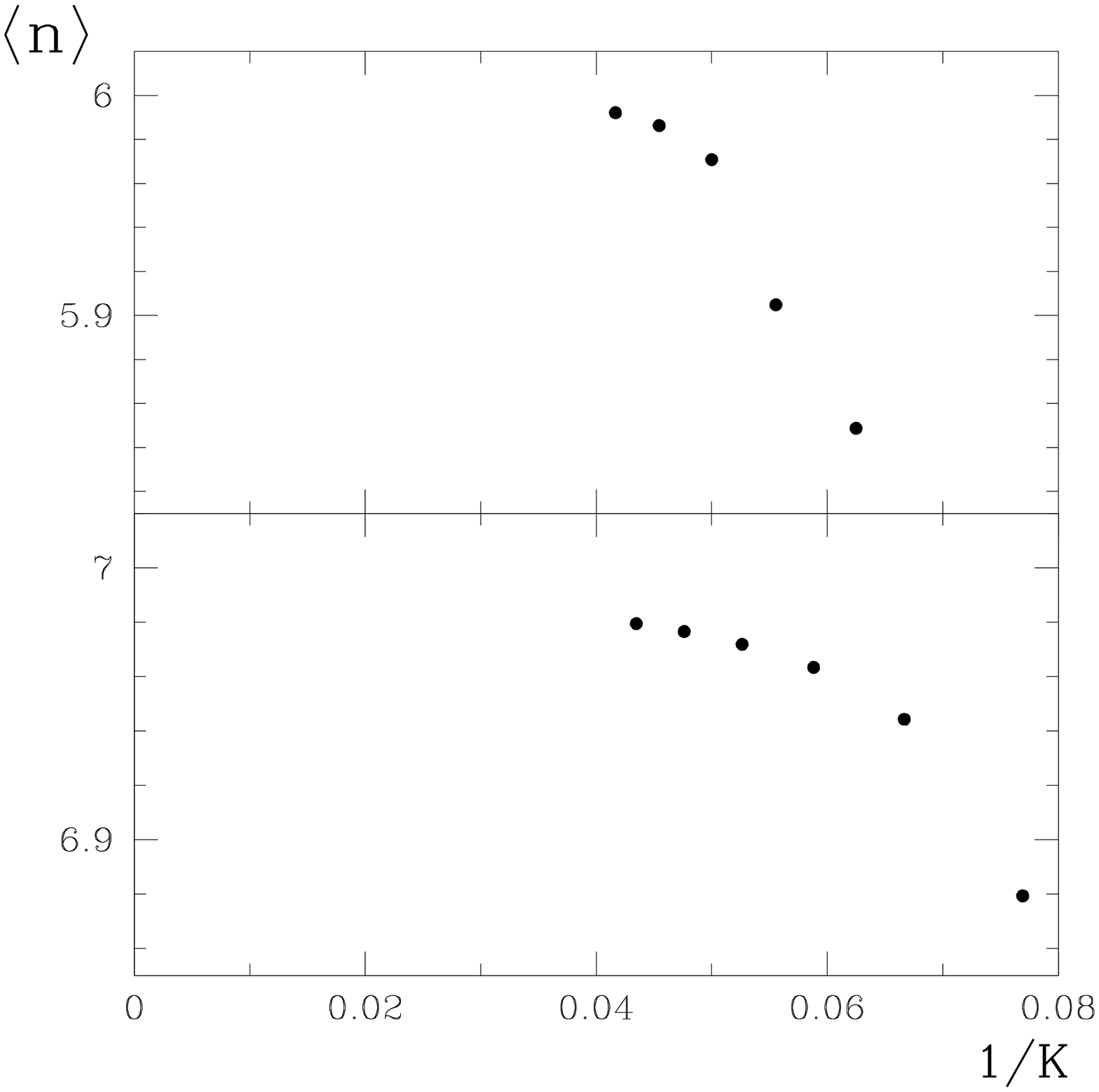,width=7.8cm}
}
\caption{
(a) Left:  
The lowest four three-parton $\cal T$-even DLCQ eigenfunctions of the theory 
with non-singular terms at $K=151$ 
(solid lines) and of the asymptotic theory (dashed lines). Of the
latter, the lowest eigenfunction has been suppressed by a factor five,
and the second(third) lowest appears as analogue of the third(fourth) 
lowest non-singular function.  
(b) Right: 
Average parton number as a function of $1/K$ of 
a $\cal T$-even boson (top, $\epsilon=0.225$, $M^2\approx 41$) and 
fermion (bottom, $\epsilon=0.505$, $M^2\approx 31$).).\label{UMfig3p}}
\end{figure}
%

In Sec.~\ref{SecNoPairs} 
we solved for the spectrum of $P^-_{asympt}\equiv P^-_{PC,s}+P^-_{ren}$
keeping only singular terms in the Hamiltonian. 
While it will be hard to find analytic
solutions without omitting non-singular operators, a numerical
solution can be obtained without any problems.
We find some noteworthy changes when regular operators are included.

The two-parton solutions are entirely unaffected by the regular terms, the 
lowest mass being ${\bar M}^2=11.74$.
In contrast, the lowest ${\cal T}$-even three-parton 
mass jumps dramatically, as the massless state acquires a 
mass (squared) of 5.703 when regular terms are present. 
Its wavefunction has the same structure as the lowest massive asymptotic
one, save for an overall shift 
due to the admixture of the constant massless wavefunction. 
We note three things. Inclusion of non-singular terms
inverts the mass hierarchy of massive states, 
namely a three-parton state becomes lighter than
a two-parton 
state\footnote{This is natural in the bosonized theory where the three
parton state corresponds to the lowest state $\Tr\{J\psi\}|0\rangle$, see 
Sec.~\ref{SecCurrents}.}. 
Secondly, this mass is very close to the continuum value 
obtain for the full theory $\bar{M}^2_{\rm full,f}=5.75$, 
{\em cf.}~$\bar{M}^2_{\rm full,b}=10.84$ of the 
lightest boson. We infer that the lowest state is 
very pure in parton
number, consistent with previous results \cite{DalleyKlebanov}.
Thirdly, the only two sectors unchanged by the inclusion of non-singular terms
are the two-parton and the ${\cal T}$-odd three parton sectors.

Why is the  ${\cal T}$-even 
three-parton sector so heavily influenced by non-singular operators?
We can get the idea by studying the wavefunctions.
In the $\cal T$-odd sector, the wavefunction is an odd function of the momenta,
see Fig.~\ref{fig3p}(a) and Eq.~(\ref{phi3minus}).
That means that the contributions from $1/(k_1+k_2)^2$ in $A(k_i)$, 
Eqn.~(\ref{EqnA}), will cancel
and the masses will stay the same. In the ${\cal T}$-even sector, 
corrections are large when the wavefunction
is large at the boundaries, {\em i.e.}~where 
(at least) one momentum vanishes, e.g.
$(0,x_2,x_3=1-x_2)$. We would therefore 
expect the first, second and fourth $\cal T$-even {\em massive} eigenvalues
to change substantially, but not the third, see the dashed wavefunctions in 
Fig.~\ref{UMfig3p}(a). 
To confirm our intuition, 
we compute the first and second corrections to the masses by sandwiching 
the operator $\langle i|2P^+P^-_{PC,r}|j \rangle$. It is easiest to do this
numerically, using the existing eigensolutions of the asymptotic (unperturbed) 
Hamiltonian. We obtain for the lowest five masses 
\bea
\bar{M}^2_0 &=& 0 + 5.961  -0.3536 =      5.607 (5.703)\nonumber \\
\bar{M}^2_1 &=& 21.59 + 8.589  -1.564 =      28.62 (29.05)\nonumber \\
\bar{M}^2_2 &=& 46.66 + 9.776  -2.136 =      54.30 (54.20)\nonumber \\
\bar{M}^2_3 &=& 56.07 + 1.662 + 0.3040 =      58.04 (59.54)\nonumber \\
\bar{M}^2_4 &=& 74.29 + 10.92  -1.562 =      83.65 (83.81), 
\eea
in agreement with our expectations. 
The non-perturbative results are listed in parentheses. 
Unsurprisingly, we find 
\[
\langle\phi_{3,i}^-|2P^+P^-_{PC,r}|\phi_{3,j}^-\rangle=0
\]
for all $i,j$, hence the corrections to the $\cal T$-odd eigenstates vanish
identically. 

Although we find a massless state in all $T=(-1)^{r+1}$ 
$r$-parton sectors, the $\cal T$-odd three-parton sector is the only one that
does not receive corrections.   
The corrections are substantial in the other sectors (4+: 40\%, $4-$: 80\%, 
5+: 57\%, $5-$: 38\%, 6+: 167\%, $6-$: 131\%; at typical resolutions $K$). 
It is remarkable that the non-singular terms generate 
most of the mass of the six-parton states.
Of course, none of this is in contradiction with the 
assumption that the asymptotic Hamiltonian is a good approximation at high
excitation numbers.

\section{The role of parton-number violating operators}
\label{SecPairs}

If we include the parton-number violating operators, we obtain the
full theory at large $N_c$. Again a numerical
solution can be obtained easily, with the caveat of a much higher number of
basis states due to the coupling of parton sectors. 
Approximate (numerical) solutions are 
well-documented in the literature, see e.g.~\cite{GHK}.

The parton-number changing interactions are three-body operators, 
and therefore have 
the largest influence on three-parton states.
The relevant function, Eq.~(\ref{B}), 
is small
when the momenta are roughly the same, and large when $k_1-k_3$ is large while
$k_2$ is small. Since $k_3=1-k_1-k_2$, the biggest contributions arise when
$k_1$ and $k_2$ are very different. 

We can investigate the role of 
parton-number violating operators using perturbation theory by parametrizing 
the Hamiltonian
\[
P^-=P^-_{asympt} +\epsilon  P^-_{PV}.
\]
Previously, we argued on 
physical grounds that $P^-_{PV}$ is marginal at high excitation
numbers without explicitly identifying a small parameter. 
Here, we use $\epsilon$ to continuously 
switch from asymptotic to full theory. 
Obviously, the only non-zero parton-blocks of $P^-_{PV}$ lie on its
upper and lower secondary diagonals. Consequently, 
there is no first-order correction to the eigenvalues.
At second order an $r$-parton eigenfunctions receives admixtures 
of $r-2$ and $r+2$ states only. In particular, 
the two(three)-parton eigenfunctions exhibit only four(five)-parton 
contaminations.

In Sec.~\ref{SecNoPairs} we found analytic expressions for a complete
set of eigenfunctions of the asymptotic Hamiltonian.
Hence, the matrix elements 
$\langle \phi_{r,\pm}|P^-_{PV}|\phi_{r+2,\pm}\rangle$ 
can in principle be calculated
analytically. However, 
it should suffice to evaluate 
the operators numerically and extrapolate to the continuum, with the
advantage of using {\em ab ovo} 
correct (at a certain $K$)  
solutions\footnote{In general, only a linear combination of the 
analytic asymptotic wavefunctions will be an eigensolution.}. 

Empirically, we find that parton-number violation is
necessary to produce (exact) 
multi-particle states in the spectrum. The reason is that only the 
complete Hamiltonian can be cast into a current-current 
form in the bosonized theory,
where the decoupling of the multi-particle states can be seen explicitly
\cite{KutasovSchwimmer}.

\begin{figure}
\centerline{
\psfig{file=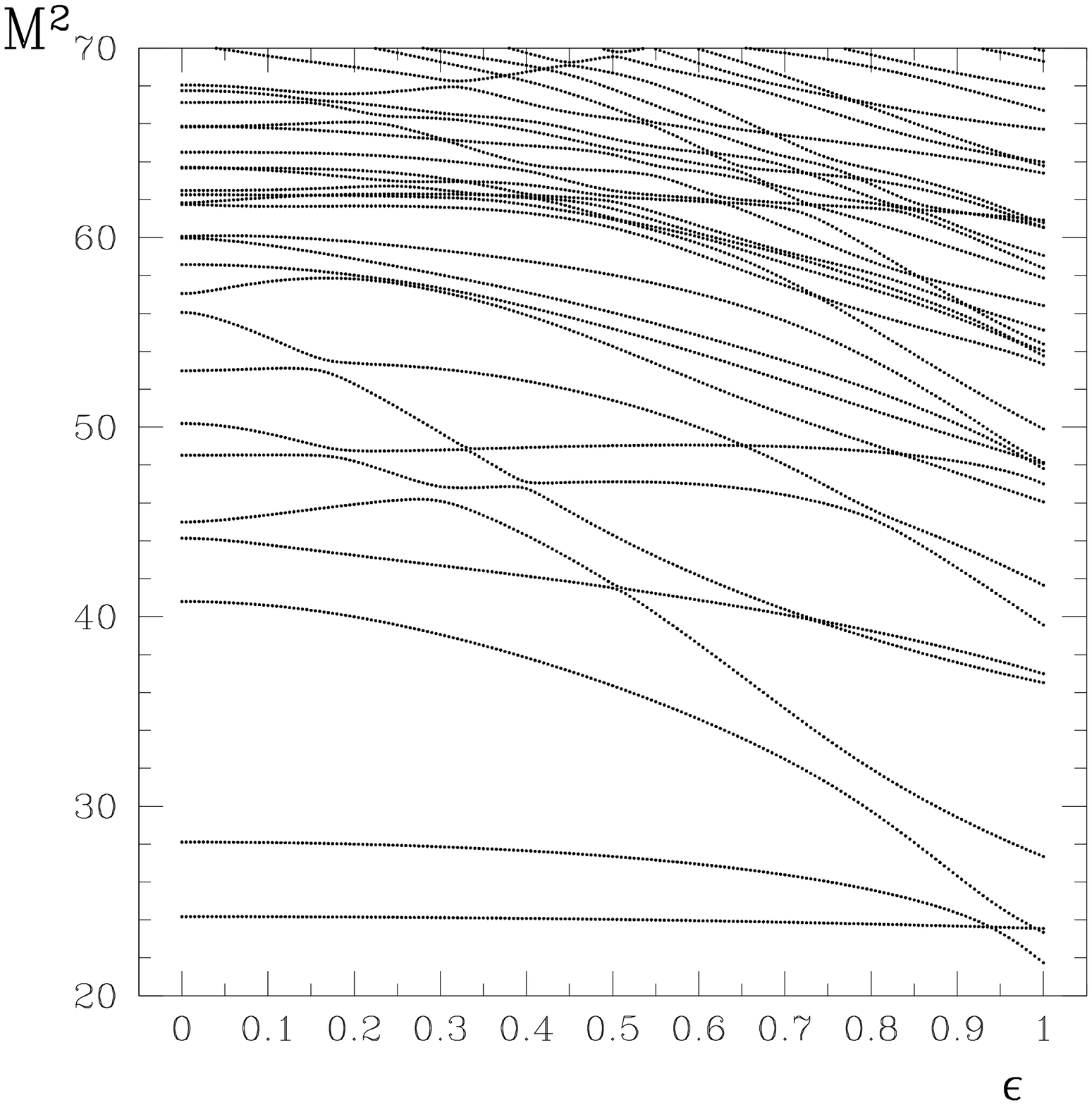,width=7.8cm}
\psfig{file=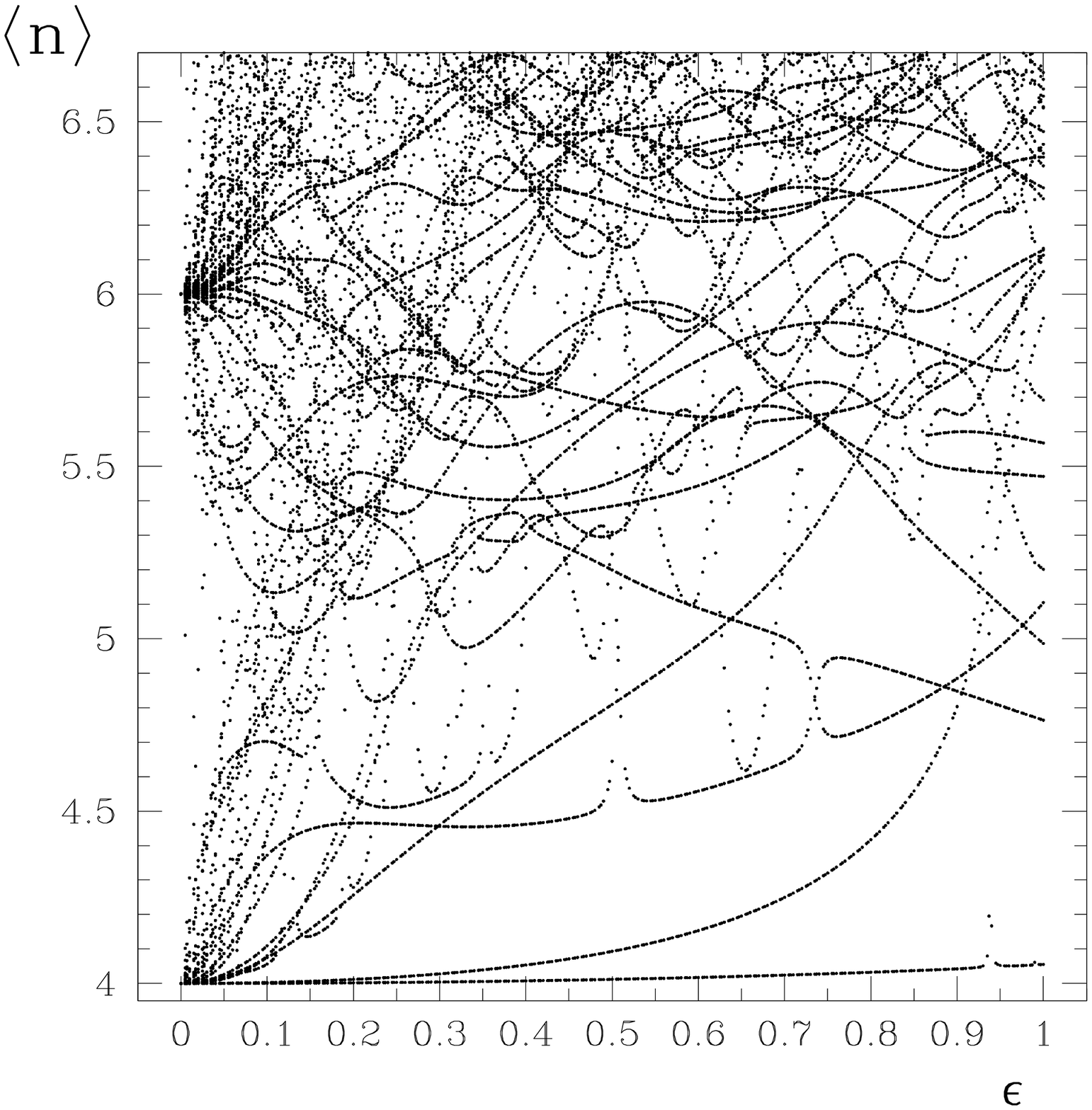,width=7.8cm}
}
\caption{(a) Spectrum (b) average parton number 
of the states in the $\cal T$-odd bosonic sector as a function of
the parton-number violation parameter $\epsilon$. 
\label{epsilonFigure}}
\end{figure}

We will use the purity of states in parton number as a measure of the 
importance of pair production. 
The authors of Refs.~\cite{DalleyKlebanov,BDK} infer that  
the lowest states of the theory are very close to being eigenstates 
of the parton-number operator. Looking at the mass versus $\epsilon$ plot, 
Fig.~\ref{epsilonFigure}(a), 
it appears that this is a by-product of the fact that the lowest
states are quite isolated in mass. Consequently, these states are mostly
inert with respect to admixtures from other parton sectors, and 
pair production is not important for the lowest states.  However, since 
there are no (exact) multi-particle states without pair production, 
it has to be crucial
for the other states. This importance may, however, not be reflected in 
parton-number impurity.

In Fig.~\ref{epsilonFigure}(a), there are two distinct behaviors when the 
masses of two states are similar,
$M_i(\epsilon)\approx M_j(\epsilon)$: either the eigenvalues repel or they 
are not influencing each other at all. To study these points, we plotted the
average number of partons $\langle n \rangle$ in a state versus $\epsilon$ in
Fig.~\ref{epsilonFigure}(b). Although distinct trajectories of several states 
are discernible, the plots give us little leverage to decide which 
are the single-particle states. Whenever two states
come close in mass, their other properties become similar, too. Although
it is interesting to observe how some states ``recover'' from mixing 
at certain values of $\epsilon$, the underlying message seems to be that
states cannot be unambiguously identified as we continuously 
turn on pair production. 
It is a little disturbing that the function 
$\langle n \rangle(\epsilon)$ of the lowest 
$\cal T$-odd boson exhibits a cusp at $\epsilon\approx 0.94$. This seems
to be a numerical artifact; the effect diminishes as $K$ grows.
   
The lowest $\cal T$-odd fermion is a very pure
five-parton state. This begs the question if the 
scheme\footnote{Lowest single-particle
states in the bosonic ${\cal T}+$, the fermionic ${\cal T}+$, 
the bosonic ${\cal T}-$, and the fermionic ${\cal T}-$ sectors 
are pure 2,3,4,5-parton states, 
respectively.} continues. We do see evidence for it. In particular, 
there is a pure 
six-parton state (up to $\epsilon\approx 0.3$ for $K=24$), 
and there is a pure 
seven-parton state for $\epsilon<0.6$. Both states are $\cal T$-even. 
We followed the development of these states at larger resolution, and they
seem to stabilize. Namely, they become purer in parton number as 
$\epsilon$ and $K$ grow, see Fig.\ref{UMfig3p}(b);
the other states in this sector become less 
pure. If we extrapolate the curves $\langle n\rangle (\epsilon,K)$ towards 
the continuum, we obtain  $\langle n\rangle=6$ and $\langle n\rangle=7$,
respectively. 
This suggests the existence of a tower of infinitely many 
single-particle states organized in a double Regge trajectory.

%
%

\section{Implications of Bosonization}
\label{SecCurrents}

The structure of the QCD$_{2A}$ spectrum is best understood 
in terms of current 
operators $J(-p)\sim\int dq\, b(q)b(p-q)$, 
{\em i.e.}~by looking at the bosonized theory \cite{Adi95}. Fermionic states 
have an additional single fermion operator. 
The eigenvalues are the same as 
in the fermionic 
picture\footnote{Although the expressions ``bosonized theory'' 
and ``fermionic picture'' appear on unequal footing, 
they help to avoid ambiguous expressions.
},
yet the eigenfunctions are not, due to the fact that bosonization is
a basis transformation. To find the single-particle states it is sufficient
to restrict calculations to the single-trace sector \cite{KutasovSchwimmer}. 
As pointed out earlier, the problem is that not 
all single-trace states are single-particle states. 
Bosonization organizes  
the single-trace sector into blocks with distinct numbers of single-fermion
operators ($f=0,1,2,\ldots$), yet 
only the blocks with $f=0,1$ give rise to 
single-particle states \cite{KutasovSchwimmer}. 
The task to rid these blocks of (approximate) 
multi-particle states to reveal the true, single-particle content of the 
theory is hard. The problem is the mixing of the 
approximate multi-particle states with the single-particle states at any
finite resolution. 
 
We can quickly confirm that the above 
block-diagonalization is realized in any framework\footnote{We describe
a DLCQ construction; analogous procedures exist whenever 
the spatial dimension is compactified.} 
with discrete momentum 
fractions. This exercise will make it easier to understand the 
role of the approximate multi-particle states by projecting out the 
exact multi-particle states. It requires 
the construction of direct-product (DP) states of the form 
\[
|DP\rangle=\Tr[J^{n_1}\psi J^{n_2}\psi \cdots J^{n_s}\psi] |0\rangle,
\]
where $J^{n_1}$ is a product of $n_1$ current operators carrying, in general, 
different (integer) momentum fractions, $\psi\equiv b(-1/2)$ is a 
fermion operator of momentum 
fraction $1/2$, and $s>1$. 
Note that by constructing the DP states, we explicitly show that in 
QCD$_{2A}$ one
cannot identify single-trace and single-particle states contrary to the 
't Hooft model \cite{tHooft}. The important result of this exercise is
that {\em only  
fermionic states} of the form $\Tr[J^{n}\psi]|0\rangle$ form exact 
multi-particle states, 
and therefore (likely) also the approximate MPS, as found numerically in 
\cite{GHK,Katz,UT}. 

The dimension of the DP sector of the 
bosonized single-trace sector plus the dimension of the (potential) 
single-particle sectors add up to the dimension of the single-trace sector
in the fermion picture, for both the fermionic and the bosonic sectors
of the theory. For instance, for $K=21/2$ one has 1169 states in the fermionic
picture, and 512 in the bosonized theory. Counting direct product 
states of the form $|K_1\rangle\otimes|K_2\rangle\otimes\cdots\otimes
|K_s\rangle$ ($\sum_j K_j=21/2$)
one arrives at 697, but 40 states of the form $[|K=7\rangle]^3$
are cyclically redundant,
see Table \ref{FockSpaceSizes}. This implies 
that cyclic permutation of ``constituent fermions'' does not 
lead to independent states, consistent with the behavior of the 
bosonized states of the bosonic sector. For example, at $K=4$, we have
\[
\Tr[J(-2)\psi J(-1)\psi]|0\rangle = \Tr[J(-1)\psi J(-2)\psi]|0\rangle,
\]  
up to terms with a lesser number of operators.
Pauli exclusion dictates that direct product states of identical 
fermionic bound states, like $\Tr[\{J(-n)\psi\}^2]|0\rangle$, 
vanish\footnote{For the general rule, see \cite{DalleyKlebanov}, Sec.~III.}. 
We note that the number of DP states implies that {\em all} states, including 
the approximate multi-particle states, form DP states, while the much smaller
number of approximate MPS (maximally the sum of the dimensions of the 
blocks with less than two single-fermion operators) 
suggests that only some, mostly likely the single-particle states, form those. 

\begin{table}[ht]
\begin{tabular}{|r|rr|rr|r||r|rr|rr|r|}\hline
& \multicolumn{2}{|c|}{Ferm.Pic.}& \multicolumn{2}{|c|}{Bos.Theory} & DP & 
& \multicolumn{2}{|c|}{Ferm.Pic.} & \multicolumn{2}{|c|}{Bos.Theory} & DP\\
$2K$ & $T+$ & $T-$ & $T+$ & $T-$ & states &
$K$ & $T+$ & $T-$ & $T+$ & $T-$ & states \\\hline
${3}{}$ & 1 & 0 & 1 & 0 & 0 & 2 & 1  & 0 & 1 & 1&0\\
${5}{}$ & 1 & 1 & 1 & 1 & 0 & 3 & 1  & 1 & 1 &  2&0\\
${7}{}$ & 3 & 1 & 3 & 1 & 0 & 4 & 4  & 2 & 3 & 1 &2\\
${9}{}$ & 4 & 5 & 4 & 4 & 1 & 5 & 5 & 6 & 3 & 3 & 5 \\
${11}{}$ & 11 & 7 & 10 & 6 & 2 & 6 & 16 & 12 & 8 & 4 & 16 \\
${13}{}$ & 18 & 22 & 16 & 16 & 8 & 7 & 27 & 31 & 9 & 9 & 40 \\
${15}{}$ & 51 & 42 & 36 & 28 & 29 & 8 & 75 & 66 & 21 & 13 & 107 \\
${17}{}$ & 99 & 111 & 64 & 64 & 82 & 9 & 153 & 165 & 29 & 29 & 260\\
${19}{}$ & 257 & 235 & 136 & 120 & 236 & 10 & 392 & 370 & 61 & 45 & 656\\
${21}{}$ & 568 & 601 & 256 & 256 & 657 & 11 & 879 & 1791 & 93 & 93 & 1605\\
${23}{}$ & 1421 & 1365 & 528 & 496 & 1762 & 12 & 2196 & 2142 & 191 & 159 & 3988\\
${25}{}$ & 3312 & 3400 & 1048 & 1048 & 4664 & 13 & 5166 & 5254 & 315 & 315 & 9790\\
${27}{}$ & 8209 & 8064 & 2080 & 2016 & 12177 & 14 & 12777 & 12632 & 622 & 558 & 24229\\
\hline
\end{tabular}
\caption{Dimension of Fock bases in the fermionic picture and 
the bosonized theory. Fermionic (bosonic) states are on the left (right). 
\label{FockSpaceSizes}}
\end{table}
  
In sum, bosonization casts approximate and exact multi-particle states into
different sectors. The hope is that this insight leads to a method to 
identify and eliminate the former from the $f=0,1$ blocks to reveal the 
true content of the theory.

\section{Conclusions}
\label{SecConclusion}

We have constructed an algebraic solution of the asymptotic approximation
to QCD$_{2A}$ in the lowest parton sectors. We were able to elucidate the 
impact of non-singular parts of the Hamiltonian on the spectrum, and presented
a perturbative calculation by smoothly turning on the parton-number violating
operators. This allowed us to present evidence for the existence of two linear 
Regge trajectories of single-particle states, in accordance with earlier 
and recent work \cite{UT, Katz}.  

While we were not able to find a criterion to distinguish single- from 
multi-particle states in general, we have presented several new facts that
can be used towards finding the single-particle spectrum of QCD$_{2A}$.
About the structure of the spectrum we learned 
the following: all states form exact multi-particle states, but
only fermionic states with exactly one fermionic operator form
approximate multi-particle states. Furthermore, coupling between
parton sectors is a necessary condition for the existence of 
multi-particle states.

Ref.~\cite{PerilsSUSY} cautions us not to read too much into differences of 
approximations to the theory at finite resolution. 
On the other hand, the appearance of 
multi-particle states has been seen in two very different approaches 
\cite{GHK, Katz}, and therefore hints at a framework-independent problem.
In classic DLCQ the Hamiltonian is block-diagonal in resolution, but the 
supersymmetry operators are not, as the additional fermion has non-zero
momentum at finite resolution. Hence, spurious interactions between
single- and multi-particle states are induced to guarantee supersymmetry
at $m_{SUSY}=g^2 N$ in the continuum limit, 
which make it hard to separate them. Even a manifestly
supersymmetric framework like SDLCQ \cite{SDLCQ} does not circumvent the
problem. The need to use periodic boundary conditions induces other
interactions, and leads to worse convergence for massless fermions. 

It may make sense to attempt to understand the spectrum 
of the theory using supersymmetry, which is exact for 
$m_{SUSY}$ and 'softly' broken otherwise \cite{BoorsteinKutasov}. 
One idea is to flesh out the 
construction of wavefunctions by applying the supersymmetry generator
sketched in 
\cite{Kutasov94}. This should work off the supersymmetric point  
\cite{PerilsSUSY} for the asymptotic theory.

\section*{Acknowledgments}
The hospitality of the Ohio State University's Physics 
Department, where most of this work was completed, is gratefully acknowledged.

\begin{appendix}


\section{Physical Hilbert Space}

To solve the eigenvalue problem, Eq.~(\ref{TheEquation}), 
we need to use a basis of the physical Hilbert space.
Due to the cyclic symmetry of states made of adjoint partons and the
fixed, total momentum set to unity, we have an integration volume in the 
$r$-parton sector\footnote{We can ignore 
states eliminated by Pauli exclusion, since they constitute a set of measure 
zero.} 
\[
\int_0^{1/r} dx_1\left(\prod^{r-1}_{i=2}
\int^{1-(r-1)x_1-\sum^{i-1}_{j=2}x_j}_{x_1}dx_i\right)=\frac{1}{r!}.
\] 
It seems that we have singled out $x_1$ and $x_r$, but the wavefunctions
are cyclic in all momentum fractions which eliminates this concern.

Our naive choice of states, the ansatz (\ref{Ansatz}), is not orthogonal
on the physical Hilbert space for $r>3$, and thus constitutes an overcomplete 
basis. However, we can find linear combinations which group the naive 
solutions into $1/r!$ conjugacy classes orthogonal on the physical 
Hilbert space. For $r<4$ we are done, because the ${\cal C,T}$ operators 
exhaust the possibilities. For $r>3$ we have to form linear combinations of
$\frac{1}{2}(r-1)!$ states.

\end{appendix}

\end{document}